# Unusual increase in electrical and thermal conductivity of severely deformed CrMnFeCoNi high-entropy alloy: Roles of disordering and nanostructuring in electron and phonon transport

Jacqueline Hidalgo-Jiménez[1], Payam Edalati[2], Md Amirul Islam[1], Makoto Arita[3], Dmitry Chernyshov[4], Bidyut Baran Saha[1], Kaveh Edalati[1,*]

[1] WPI, International Institute for Carbon Neutral Energy Research (WPI-I2CNER), Kyushu University, Fukuoka 819-0395, Japan

[2] Faculdade de Engenharia Mecânica (FEM), Universidade Estadual de Campinas (UNICAMP), Limeira, Brazil

[3] Department of Materials, Faculty of Engineering, Kyushu University, Fukuoka 819-0395, Japan

[4] Swiss-Norwegian Beam Lines, European Synchrotron Radiation Facility, 38000 Grenoble, France

This work investigates how defect engineering and nanostructuring via severe plastic deformation influence electrical and thermal transport in the high-entropy alloy CrMnFeCoNi. High-pressure torsion (HPT) for up to $N = 4$ turns produces a hierarchical defect structure, including deformation twins at low strains, dislocations and Lomer-Cottrell locks at intermediate strains, and a high fraction of high-angle grain boundaries at large strains. While hardness naturally increases and saturates at ~520 Hv, both electrical and thermal conductivity exhibit an unusual increase upon HPT processing (from 0.953 MS $m^{-1}$ and 10.3 W $m^{-1}$ $K^{-1}$ after homogenization to 1.003 ± 0.005 MS $m^{-1}$ and 14.4 W $m^{-1}$ $K^{-1}$ after $N = 4$). Analysis based on the Wiedemann-Franz law indicates a modest increase in electron transport (~5%) and a significant increase in phonon transport (~100%) after HPT processing. The phonon mean free path remains small on the order of 0.6-1.1 nm, and comparison with the Cahill-Pohl model shows that heat transport occurs near the minimum thermal conductivity limit, where vibrational modes are strongly scattered. This suggests that the enhancement in thermal transport is not governed by conventional defect scattering, but it is hypothesized that modifications in atomic-scale disorder, such as deformation-induced short-range ordering, are responsible for this unusual behavior. Specific heat capacity increases after HPT, consistent with stored defect energy. The linear thermal expansion exhibits a non-monotonic behavior, decreasing initially due to internal stresses and relaxation effects, and increasing at higher strains due to grain boundary-induced anharmonicity. These findings highlight the complex interplay between defects, atomic-scale disorder, and nanostructure in governing the electrical and thermal properties of severely deformed high-entropy alloys via electron and phonon scattering.

* Corresponding author: K. Edalati (E-mail: kaveh.edalati@kyudai.jp; Tel: +81-92-802-6744)

## 1. Introduction

High-entropy alloys (HEAs), specifically the equiatomic CrMnFeCoNi Cantor alloy with face-centered cubic (FCC) structure, have received significant attention because of their exceptional properties, such as high strength-ductility synergy, fracture toughness and resistance to corrosion [1,2]. These properties arise from their chemically disordered lattice, intensive structural distortion, cocktail effect and sluggish atomic diffusion kinetics, which hinder dislocation motion and grain boundary migration [3,4]. Increasing the hardness and strength of HEAs has been a matter of research for at least two decades [3,4], and various strategies, including severe plastic deformation (SPD) [5,6], have been utilized for strengthening these materials. While the mechanical behavior of HEAs treated by SPD methods, like high-pressure torsion (HPT) [7,8], has been extensively studied [9-12], their functional properties, particularly electrical and thermal properties, remain largely unexplored.

SPD techniques like HPT are known to refine grain sizes to the nanometer regime and introduce high dislocation densities, which can drastically alter transport properties [5,6]. In conventional metals, defect generation and grain refinement typically increase electrical resistivity, i.e. reduce the electrical conductivity, due to enhanced electron scattering at boundaries and defects, while thermal conductivity also decreases due to additional phonon-boundary scattering [13-15]. However, the chemically complex nature of HEAs, with their inherent lattice distortion and unique defect energetics [3,4], may lead to deviations from these trends. For instance, the high entropy of mixing can modify electron-phonon coupling and defect scattering mechanisms, potentially leading to unconventional conductivity behavior [16,17]. Despite these implications, no systematic study has examined the electrical and thermal conductivity of SPD-processed HEAs, leaving a critical gap in understanding their functional properties under extreme microstructural refinement.

There have been a few efforts in past years, particularly by the group of current authors, to examine functional properties of HPT-processed HEAs [18]. For example, in the Cantor alloy, it was demonstrated that dislocation substructures at intermediate strains enhance strength while maintaining hydrogen embrittlement resistance, whereas nanograins at high strains lead to catastrophic failure due to hydrogen-enhanced decohesion [19]. Another study reported that HPT-induced nanostructuring in TiZrHfNbTa enhances superconducting properties by increasing vortex pinning sites [20]. It was also found that the HPT process can enhance the resistance to corrosion [21], improve biocompatibility [22] and hydrogen storage properties [23] of HEAs. There have also been attempts to examine the photocatalytic activity of HPT-synthesized materials for hydrogen production [24], methanation [25] and antibiotic degradation [26]. However, a survey of the literature indicates that analogous investigations for electrical and heat transport properties of severely deformed HEAs are absent. Moreover, other thermal properties of these materials, including heat capacity and linear thermal expansion, have not been investigated. Given the growing interest in HEAs for some specific functional applications, such as thermoelectric materials [27], superconductors [28], high-temperature coatings [29] and refractory materials [30], understanding their electrical and thermal properties after severe straining is essential.

Anomalous electrical transport in deformed CrMnFeCoNi has been reported [31], where it was shown that cold rolling destroys mesoscopic chemically modulated domains, decreasing electrical resistivity via the K-effect, with chromium and manganese implicated in these domains. This confirms that chemical-order changes can dominate resistivity under plastic deformation, but the effect of intense plastic deformation on transport phenomena remains limited [32]. The present work thus focuses on a different processing route and scope: (i) HPT is used, which produces a hierarchical defect structure of twins, Lomer–Cottrell locks, dislocation cells, and nanograins, and (ii) thermal transport is additionally measured to separate electronic and phononic contributions.

This study, thus, systematically explores the impact of HPT-induced lattice defects and nanostructuring on the electrical and thermal conductivities, heat capacity and linear thermal expansion of the CrMnFeCoNi Cantor alloy. Nanostructural features and lattice defects, examined via X-ray diffraction (XRD) and transmission electron microscopy (TEM), are linked with hardness and thermoelectrical properties across a range of strains. The findings reveal anomalous property-microstructure relationships for transport phenomena in HEAs, advancing understanding of defect-dominated transport in chemically complex systems that differ from conventional metals and alloys.

## 2. Experimental Procedures

The equiatomic CrMnFeCoNi alloy was synthesized through arc melting of pure (>99.9%) elements under an argon atmosphere. Discs with a 5 mm radius and 0.85 mm thickness were cut using electrical discharge machining (EDM) from the ingot, and homogenized at 1373 K for 60 min and rapidly cooled down in iced water to ensure chemical uniformity (mean grain size: 30 μm). The homogenized samples (referred to as $N = 0$ in figures) were treated using HPT under 6

GPa at room temperature, with a rotation speed of 1 rpm for $N$ = 1/16, 1/8, 1/4, 1 and 4 rotations. The imposed shear strain, $\gamma$ was estimated as [7,8].

$$\gamma = 2\pi rN/h \tag{1}$$

where $r$ is the radial distance on the disc, and $h$ is the disc height. The samples were investigated by different characterization methods, as explained below.

For the first step of characterization of materials, XRD was carried out by Cu-Kα radiation (40 kV and 40 mA) with a scanning step of 0.05° over $2\theta$ = 20-100° diffraction angle.

The samples were then examined by TEM, for which thin foils were prepared from the disc edges by cutting 3 mm diameter discs, mechanical polishing to 0.1 mm thickness, twin-jet electropolishing for electron transparency (5% perchloric acid in methanol at 263 K and 12 V) and ion milling for final cleaning (argon ions at 5 kV and incident angles of ±5º). Bright-field (BF), dark-field (DF), selected-area electron diffraction (SAED), high-resolution (HR) and fast Fourier transform (FFT) analyses were conducted at 200 kV. Grain sizes were estimated utilizing DF images by measuring the average size of bright regions.

Vickers microhardness (in Hv) was measured at 0.15, 1, 2, 3, and 4 mm away from the center of discs (500 gf load and 15 s dwell time) after polishing the samples to mirror-like surfaces.

Electrical conductivity was measured at room temperature using a four-point probe set-up. Bars with a cross-section of $A$ = 0.6×0.6 mm$^2$ and a length of 9 mm were prepared using EDM from the discs, with the center of every bar located at a fixed radial distance of 2 mm from the disc center. Prior to testing, the bar surfaces were mechanically polished to ensure reliable electrical contact, and spring-loaded pins were used to maintain contact during measurement. A direct current of $I$ = 0.2 A was passed through the two outer electrodes, and the potential difference $V$ between the two inner electrodes, separated by a distance $l$ = 4 mm, was recorded. The resistance was obtained as $R = V/I$, and the electrical conductivity of the central region of each bar was calculated as $\sigma = l/(RA)$. The bar dimensions were measured with a digital micrometer at several positions, and the mean values were used. For every condition, three independent measurements were performed and the standard deviation is reported as the uncertainty.

Specific heat capacity was obtained with a specified heat-flux differential scanning calorimeter (DSC) operated in the heat-capacity mode; the instrument has a baseline noise below 0.5 µW and a manufacturer-specified limit of permissible relative error of ±5%. Specimens weighing about 30 mg were cut from the HPT discs and scanned together with a reference pan at a heating rate of 5 K $min^{-1}$ over the selected temperature interval of 293-353 K. The heat-flow difference between the specimen and reference was recorded continuously, and $C_P$ was automatically evaluated from the calibrated heat-flow signal of the machine. Each condition was measured at least five times, and the relative standard deviation remained below 5%.

The thermal diffusivity was measured by a laser flash apparatus. In this method, a short pulse of laser is applied to one surface of the specimen, and the resulting transient temperature rise on the opposite surface is measured using an infrared sensor to estimate the thermal diffusivity ($\alpha$). Four measurements in the temperature range of 317-330 K indicated an error bar below 2%. To calculate the thermal conductivity, experimental density measured using the Archimedes method ($\rho$ = 7904 ± 20 kg $m^{-3}$) and experimental specific heat capacity ($C_P$) measured using DSC were used [33,34].

$$\kappa = \alpha\rho C_P \tag{2}$$

The thermal conductivity was determined within a temperature range of 300-330 K using one sample and four measurements at different temperatures. The relative uncertainty in $\kappa$ was estimated by propagating the independent contributions according to $\Delta\kappa/\kappa = [(\Delta\alpha/\alpha)^2 + (\Delta\rho/\rho)^2 + (\Delta C_P/C_P)^2]^{1/2}$. Using $\Delta\alpha/\alpha < 2\%$, $\Delta\rho/\rho < 1\%$ and $\Delta C_P/C_P < 5\%$, the combined relative uncertainty was about 5.5%.

For linear thermal expansion measurements, samples were measured at the BM01 beamline (Swiss-Norwegian Beamlines at the European Synchrotron Radiation Facility, Grenoble, France) with a wavelength of 0.071813 nm and using Pylatus@SNBL diffractometer [35] with 2D Pilatus3 2M detector (dectris.com). The samples were cut and thinned to a size of almost ~0.3×0.6×0.6 $mm^3$, which can cover the whole beam size of 100×300 $\mu m^2$. Calibration of the instrument was done with a standard $LaB_6$ sample using pyFAI software [36]. The Debye-Sherrer powder diffraction rings were gathered at a temperature range of 150-500 K and processed by azimuthal integration with BUBBLE software [35], as shown in Fig. 1. The temperature of the sample was controlled with a Cryostream 700+ nitrogen blower. Unit cell dimensions were refined assuming the $Fm\bar{3}m$ space group and averaged composition with FullProf software [37], and thermal expansivity was examined with the PASCAL web tool [38].

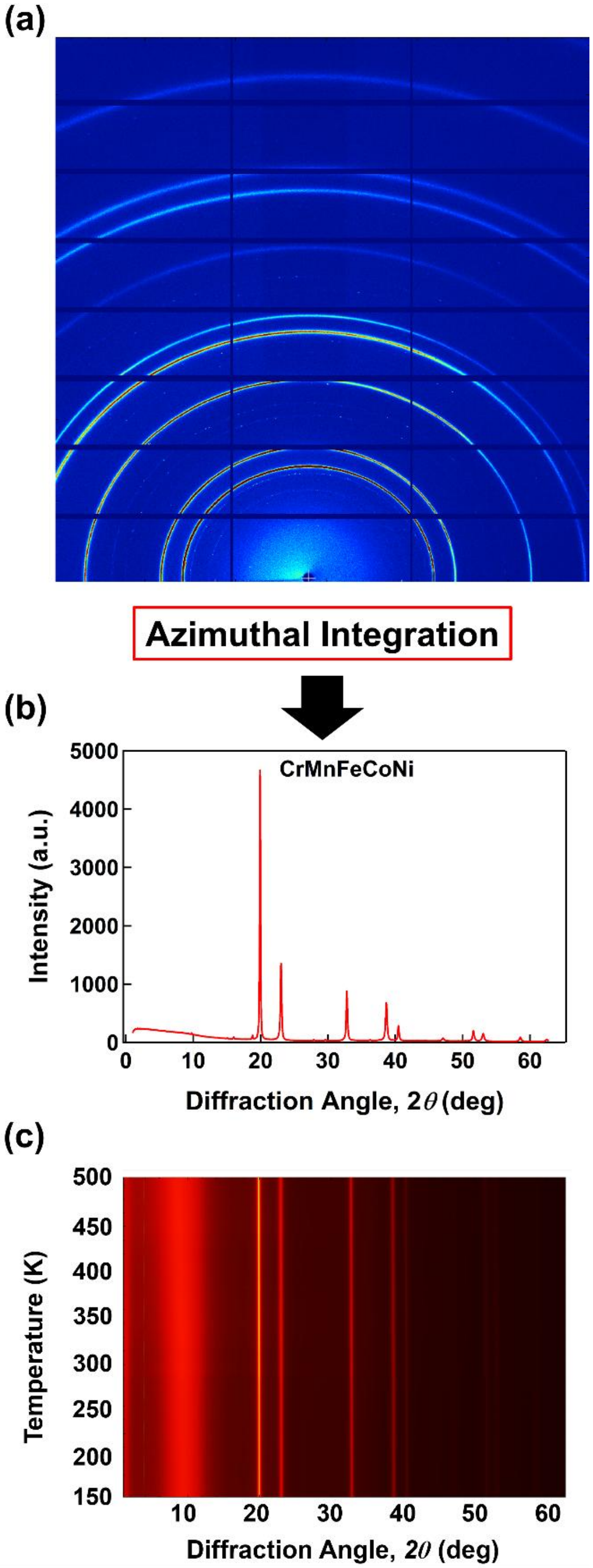

Fig. 1. Examples of measurement in synchrotron beamline for thermal conductivity measurement. (a) Debye-Sherrer powder diffraction rings and corresponding (b) diffraction profiles achieved by azimuthal integration, and (c) diffraction profiles recorded in temperature range of 150-500 K.

## 3. Results

### 3.1. Lattice Strain

The structural evolution of the CrMnFeCoNi alloy under HPT processing was systematically characterized using XRD. Fig. 2a presents the XRD patterns of samples processed with varying numbers of HPT turns. The homogenized material exhibits sharp diffraction peaks corresponding to the FCC crystal structure, with the (111), (200), and (220) peaks clearly visible at their expected positions. No phase transformations are detected, and the lattice parameter also remains unchanged at 0.3598 ± 0.0001 nm. As the HPT rotations increase, several important changes become apparent in the XRD patterns. The peaks gradually broaden, with the full width at half maximum (FWHM) of the (111) peak increasing from 0.15° in the homogenized state to 0.52° after $N = 4$ turns, as shown in Fig. 2b. This peak broadening results from two primary factors: reduction in crystallite size and introduction of crystal lattice defects. Such peak broadening, which is indicative of lattice strain and residual stress, is a common feature of severely deformed materials, including HEAs [12,20].

### 3.2. Hardness

The mechanical behavior of the HPT-treated samples, as an indicator of lattice defect evolution, was characterized through microhardness measurements. Fig. 2c presents the hardness variation across the diameter of the HPT-processed discs. In the early deformation stages ($N = 1/16$ turns), the hardness shows significant radial dependence, with values increasing from about 200 Hv at the center to 300 Hv near the edge. This gradient reflects the strain distribution inherent to the HPT technique, where the shear strain is proportional to the radial position [5-8]. As the number of turns increases, this gradient gradually disappears, and by $N = 4$ turns, the hardness becomes essentially uniform across the disc at approximately 520 Hv. The effect of HPT can be shown more evidently by examining the relationship between hardness and applied shear strain, as shown in Fig. 2d. Three distinct regimes can be identified in this plot. In the initial stage (shear strain $\gamma < 5$), the hardness increases rapidly with a slope of about 80 Hv per unit strain. The second regime ($5 < \gamma < 20$) shows a reduced hardening rate of approximately 20 Hv per unit strain. Finally, at the highest strains ($\gamma > 20$), the hardness reaches a plateau, where further deformation does not introduce additional strengthening. This appearance of steady state for hardness, which was also reported for tensile strength [19], does not necessarily mean the saturation of microstructure, but it appears to be usually a result of a balance between the strengthening and softening events via dynamic recovery/recrystallization and grain boundary migration [5-8].

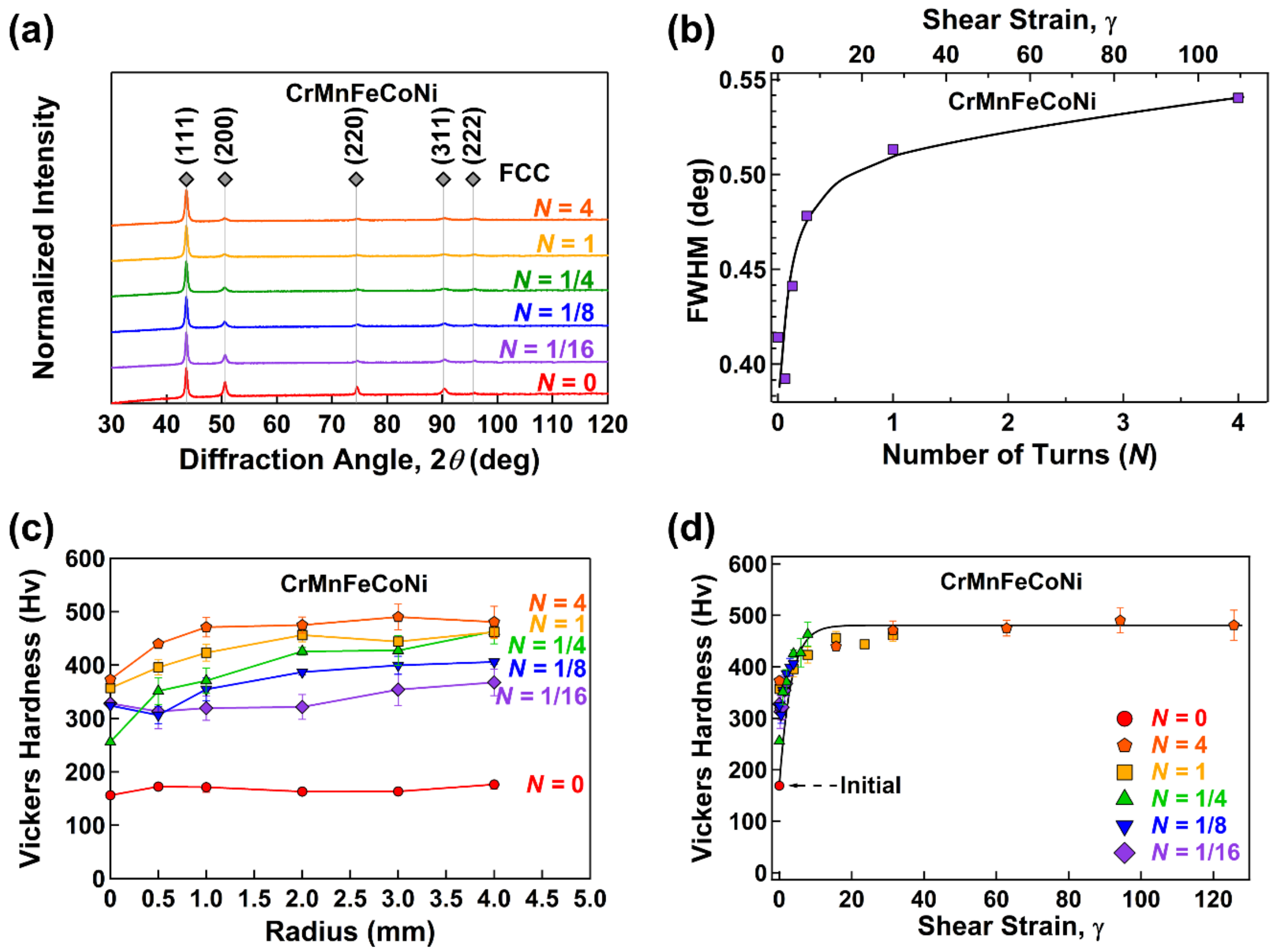

Fig. 2. XRD peak broadening and hardness increase in high-entropy alloy CrMnFeCoNi processed via severe plastic deformation. (a) XRD profiles, (b) FWHM of (111) peak versus number of rotations, (c) hardness versus radial distance in discs, and (d) hardness versus shear strain for discs processed by homogenization and HPT treatment for various rotations.

### 3.3. Microstructure

Fig. 3-5 presents a comprehensive TEM analysis of the microstructural evolution for different samples, as discussed below. For analyzing all TEM images, SAED patterns and FFT diffractograms, a lattice parameter of 0.3598 ± 0.0001 nm was used based on XRD results, while no secondary phase was detected in SAED and FFT profiles. It should be noted that attempts to examine these samples by electron backscatter diffraction (EBSD) analysis were not successful due to extremely refined and distorted microstructures.

The homogenized material (Fig. 3a) consists of large grains, with an SAED pattern showing distinct spots, confirming the single-crystal nature of these large grains.

After $N$ = 1/16 turns (Fig. 3b), the microstructure undergoes significant changes, with the deformation twins becoming the predominant feature. The SAED pattern in Fig. 3b now shows additional spots corresponding to two sets of twinned regions, maintaining a 60° orientation relationship with each other and with the matrix. This twinning activity is characteristic of materials with low stacking fault energy [39], like the Cantor alloy when subjected to plastic deformation at relatively low strains [40].

At $N$ = 1/8 turns (Fig. 3c), the deformation mechanism apparently transits from twinning to dislocation-mediated plasticity. The TEM BF and DF images reveal a high density of dislocations organized into cellular structures. TEM HR images for the disc treated with $N$ = 1/8 turns, shown in Fig. 4a, provide atomic-scale details of these defects. Numerous individual dislocations (Fig. 4b) and Lomer-Cottrell locks (Fig. 4c) are generated by the interaction of dislocations on different slip systems. These sessile dislocations create strong local strain fields, as evidenced by the distortion of lattice fringes. It should be noted that the presence of dislocations and Lomer-Cottrell locks does not necessarily mean that they cause delayed twinning.

Further deformation to $N$ = 1/4 (Fig. 3d) leads to the generation of ultrafine grains with sizes around 100 nm. The SAED pattern begins to develop a ring-like appearance, indicating the existence of multiple grains with random misorientations in the examined area. The grain boundaries at this stage should contain a substantial fraction of low-angle boundaries, as suggested by the persistence of some diffused spot-like features in the diffraction pattern.

The samples heavily deformed by $N$ = 1 and 4 (Fig. 3e and 3f) exhibit a fully nanocrystalline structure. The grain size stabilizes after $N$ = 4, reaching approximately 25 nm, and SAED patterns show complete rings, confirming the random orientation distribution of these nanograins. TEM HR, illustrated in Fig. 5, confirms the presence of high-angle grain boundaries (Fig. 5b), although some dislocation cells still remain (Fig. 5c). Interestingly, TEM lattice images reveal that individual dislocations still persist within these nanograins (Fig. 5d and 5e). The presence of individual dislocations within nanograins contrasts with the conventional expectation that grain boundaries should function both as dislocation sources and as perfect sinks for dislocations [41,42]. This rather unusual observation can be attributed to the unique chemical complexity of HEAs, where the existence of numerous principal elements with diverse atomic sizes and bonding characteristics significantly hinders dislocation motion and annihilation [1-4].

These microstructural examinations suggest that the first hardening region in Fig. 2d corresponds to the twinning-dominated deformation regime, the second region corresponds to a transition to dislocation-mediated plasticity, and the steady-state region corresponds to the presence of nanograins. Such microstructural transitions and grain refinement have been reported in a wide range of HEAs after SPD processing [43-45].

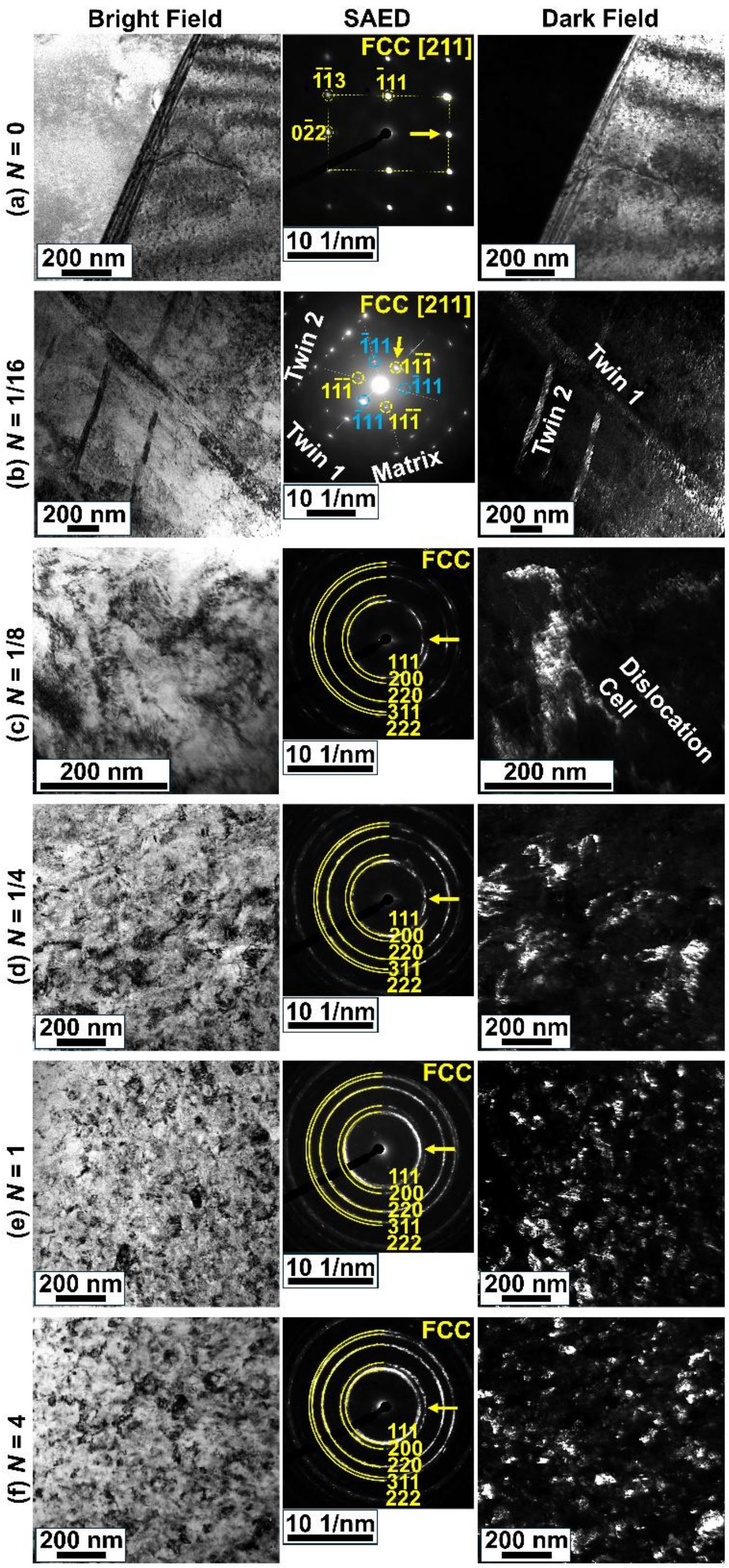

Fig. 3. Transition from coarse grains to nanograins through deformation twinning (low strains) and dislocations (intermediate strains) in high-entropy alloy CrMnFeCoNi treated via severe plastic deformation. TEM BF, SAED and DF images for discs treated by (a) homogenization and (b-f) HPT for $N = 1/16\text{-}4$ rotations.

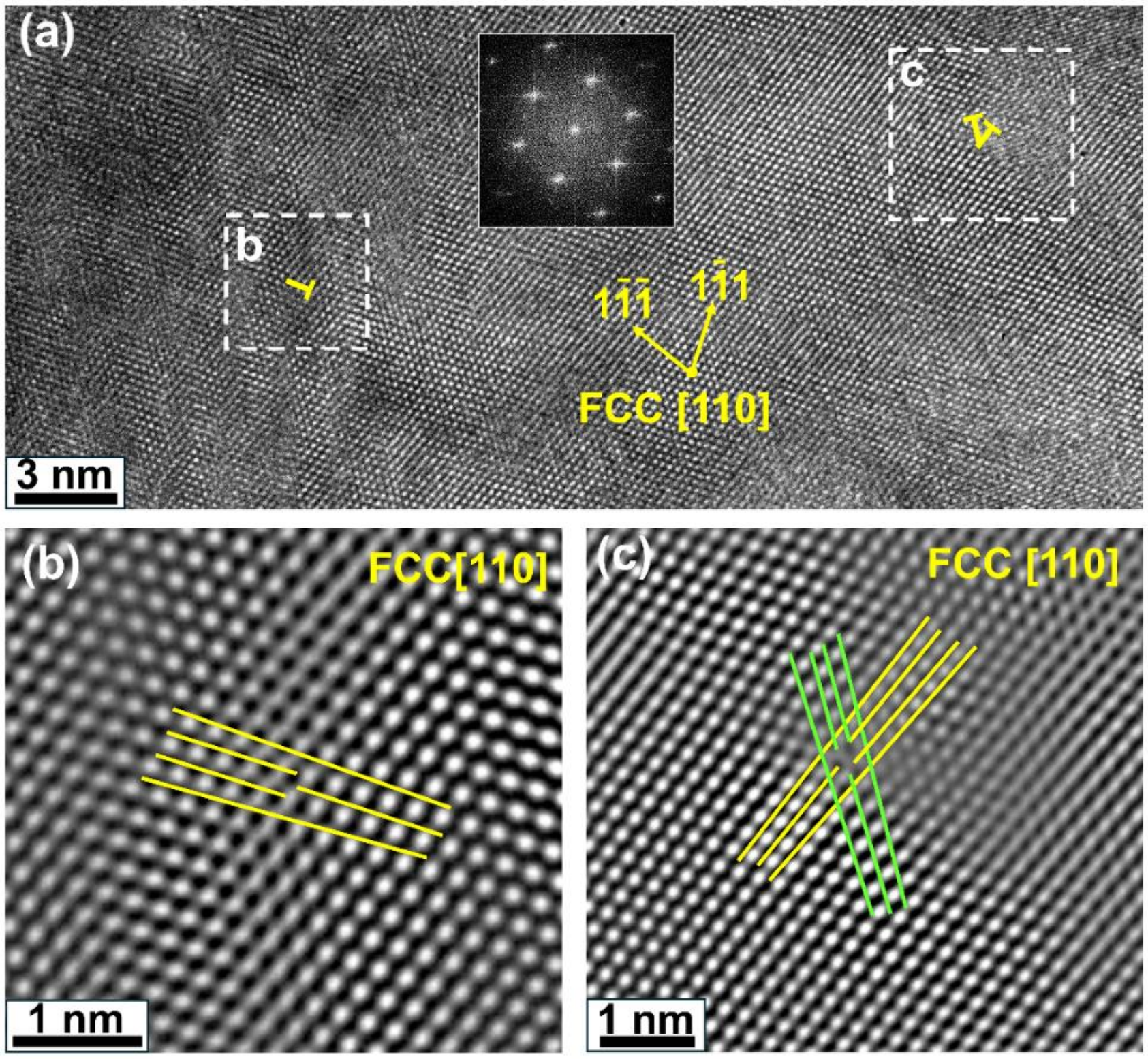

Fig. 4. Dislocation activity in high-entropy alloy CrMnFeCoNi treated via severe plastic deformation at intermediate strain ranges. (a) TEM HR image and corresponding FFT, (b) lattice image of dislocation constructed by inverse FFT analysis of area indicated in (a) and (c) lattice image of Lomer-Cottrell lock constructed by inverse FFT of area indicated in (a) for disc treated with HPT for $N = 1/8$ rotations.

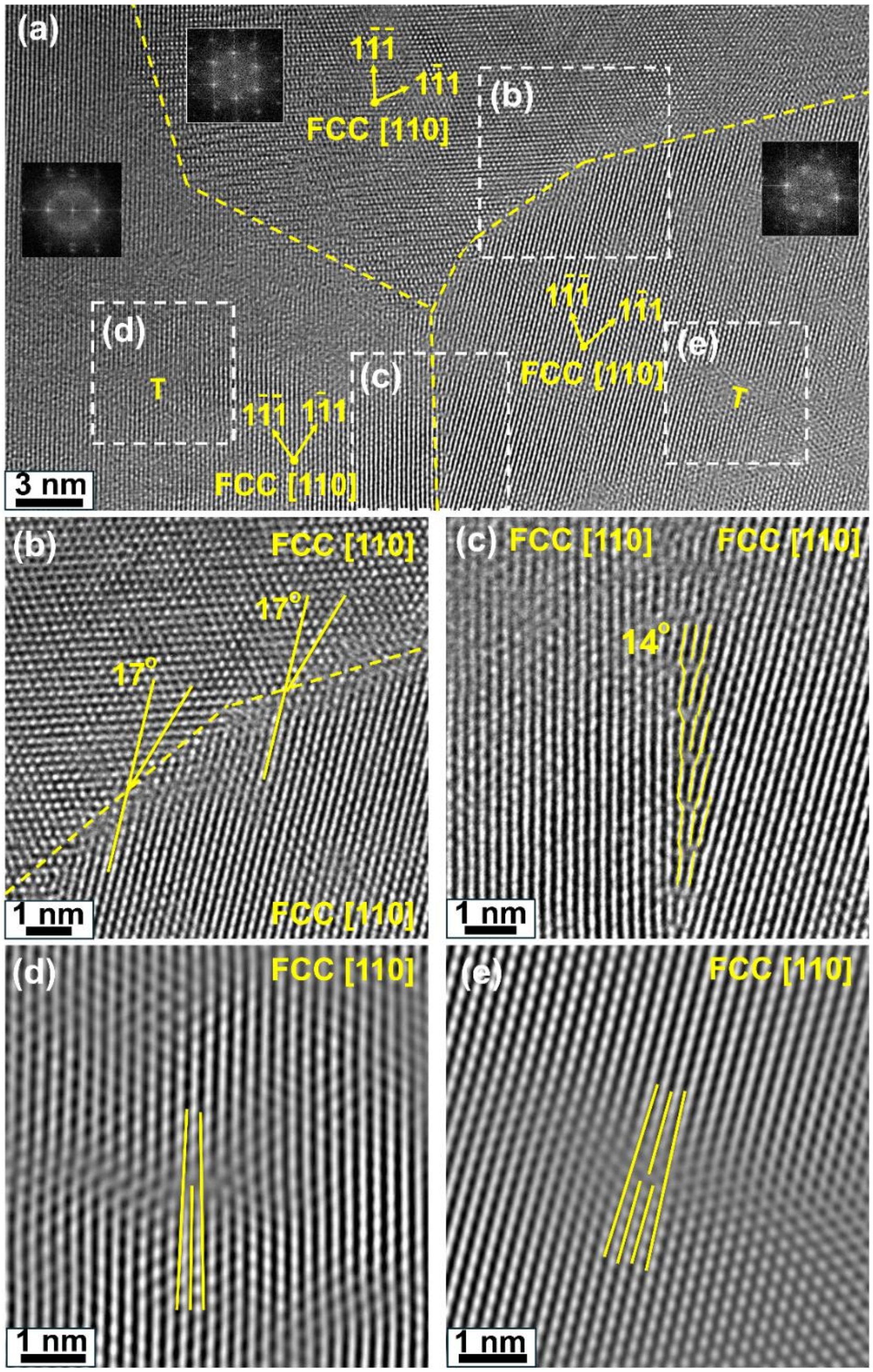

Fig. 5. Presence of high-angle grain boundaries, dislocation cells and individual dislocations in high-entropy alloy CrMnFeCoNi treated via severe plastic deformation at large strain ranges. (a) TEM HR image and corresponding FFT for three grains, (b) lattice image of high-angle grain boundary indicated in (a), (c) lattice image of dislocation cell indicated in (a) and (d, e) lattice images of dislocation constructed by inverse FFT of regions indicated in (a) for disc treated with HPT for $N = 4$ rotations.

### 3.4. Electrical Conductivity

Electrical conductivity measurements in Fig. 6 reveal a systematic increase with deformation. In this plot, three independent measurements were conducted for each HPT condition, whose differences are reflected as error bars. The homogenized material has an electrical conductivity of 0.953 ±0.004 MS $m^{-1}$, which agrees with the reported results in the literature [46]. Electrical conductivity rises gradually with increasing HPT rotations, and after a maximum at $N$ = 1, it slightly decreases and reaches 1.003 ± 0.005 MS $m^{-1}$ after $N$ = 4 turns (5% higher than the homogenized sample). Since the decrease in conductivity from the $N$ = 1 maximum to $N$ = 4 is small and close to the combined measurement uncertainty, shown by error bars, the manuscript does not assign strong physical significance to it. The increase in electrical conductivity by HPT processing is an unusual behavior, as conventional severely deformed materials exhibit a reduction in electrical conductivity by increasing applied strain due to enhanced electron-defect scattering [47,48]. This issue needs to be discussed in connection with thermal conductivity results because electron scattering plays a vital role in thermal conductivity as well [33].

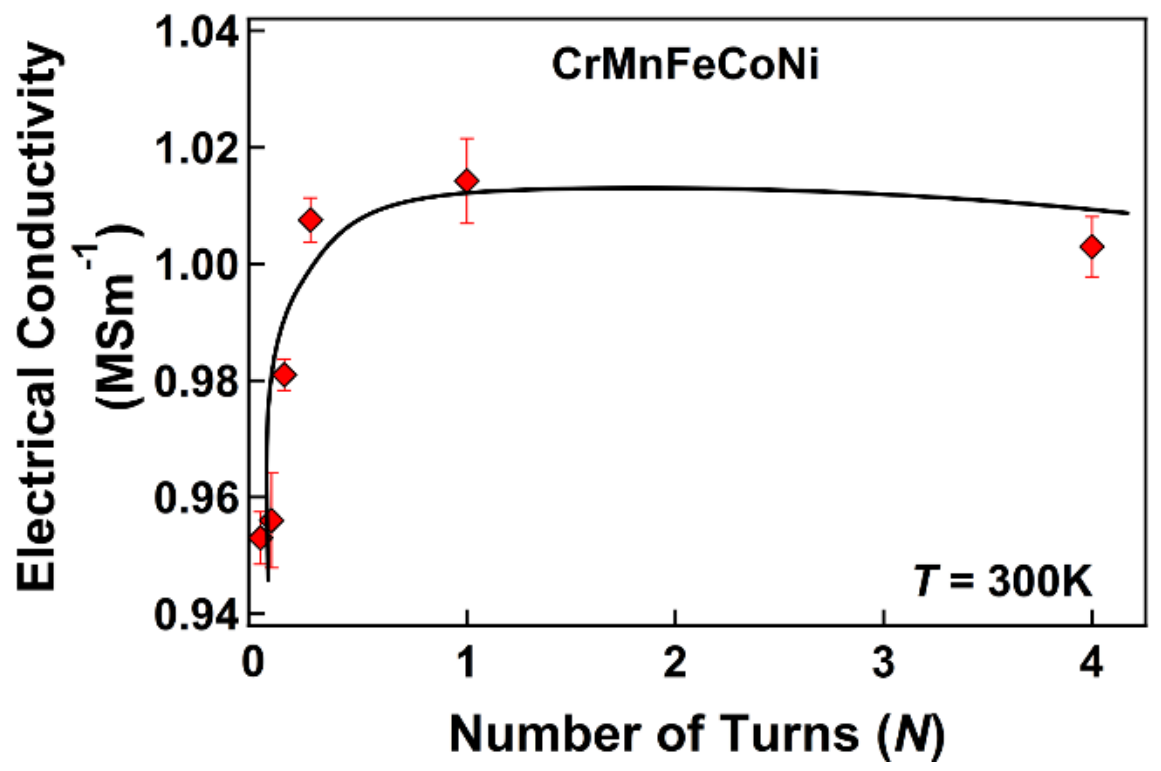

Fig. 6. Anomalous increase in electrical conductivity in high-entropy alloy CrMnFeCoNi treated via severe plastic deformation. Electrical conductivity versus number of HPT rotations.

### 3.5. Heat Capacity and Thermal Conductivity

Variations of specific heat capacity versus temperature for discs treated by different HPT turns are presented in Fig. 7a. Heat capacity increases with increasing temperature because more vibration modes of phonons become thermally activated. Another feature of Fig. 7a, which is shown more clearly in Fig. 7b, is the enhancement of heat capacity with increasing HPT rotations.

This feature originates from the capability of crystal lattice defects and grain boundaries to store energy [49,50].

The thermal conductivity results versus temperature (Fig. 7c) present an expected behavior with temperature: the conductivity increases with increasing temperature due to the activation of vibrational modes. However, the variation of thermal conductivity versus the number of HPT rotations (Fig. 7d) exhibits an unusual enhancement of conductivity after HPT processing. The homogenized material has an average thermal conductivity of 10.3 W $m^{-1}$ $K^{-1}$ within the temperature range of 317-320 K, in good agreement with the presented conductivity values in the literature [46]. The thermal conductivity increases to 14.4 W $m^{-1}$ $K^{-1}$ after $N$ = 4 turns. This behavior is unexpected, considering the effect of lattice defects on enhancing electron and phonon scattering [51,52], although electrical conductivity measurements also suggest that electron scattering is not enhanced in this HEA by HPT processing.

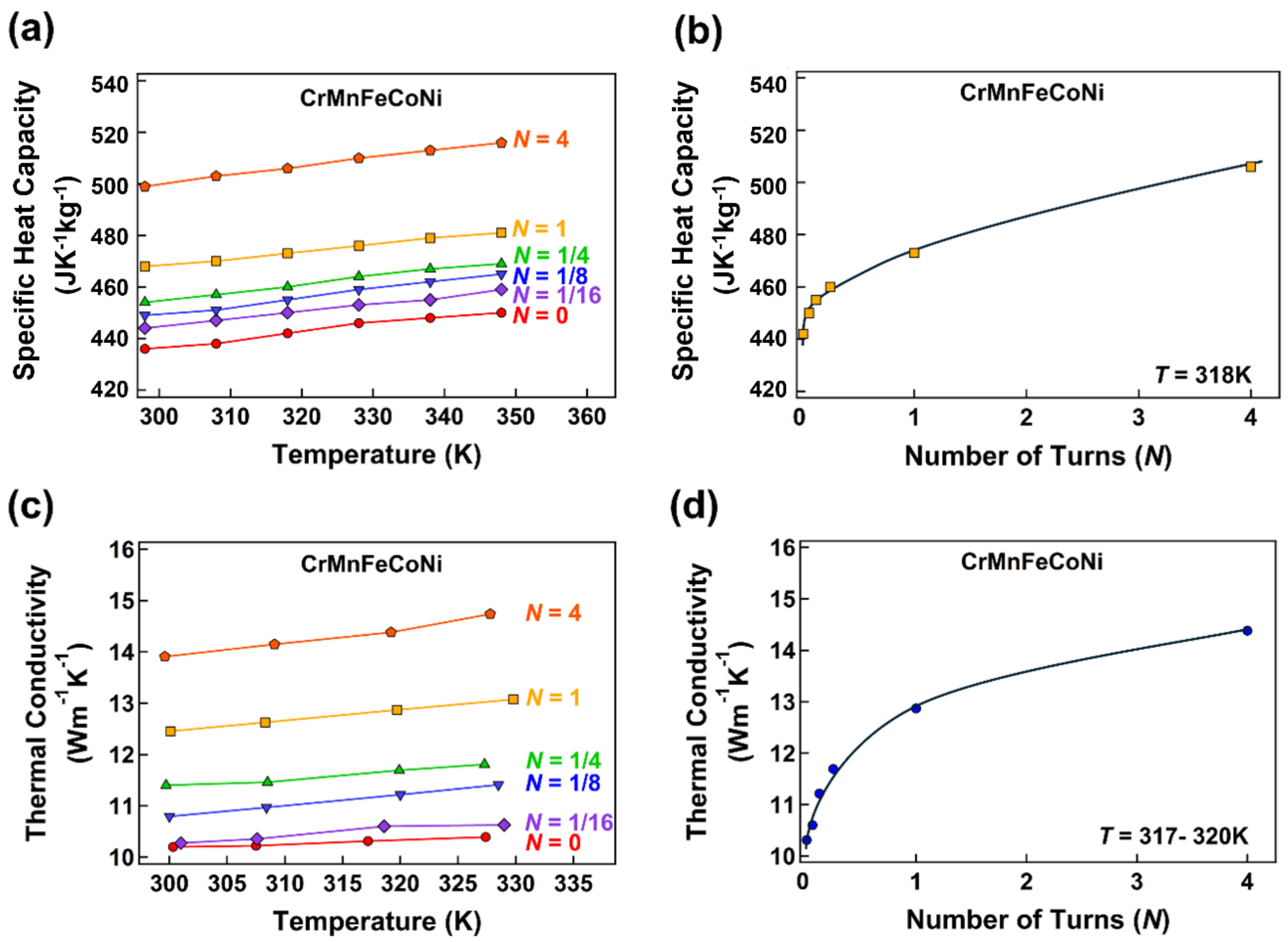

Fig. 7. Increase in heat capacity and thermal conductivity of high-entropy alloy CrMnFeCoNi treated via severe plastic deformation. (a) Specific heat capacity against temperature, (b) specific heat capacity against number of rotations, (c) thermal conductivity against temperature and (d) thermal conductivity against number of HPT rotations for discs treated by homogenization and HPT for $N$ = 1/16-4 rotations. Average thermal diffusivities for four different measurements in (c) are 2.94, 2.99, 3,12, 3.22, 3.44 and 3.60 mm$^2$ s$^{-1}$ for $N$ = 0, 1/16, 1/8, 1/4, 1 and 4, respectively, with error bars smaller than 2%.

### 3.5. Thermal Expansion

The evolution of the linear thermal expansion coefficient with HPT processing, as presented in Fig. 8, exhibits a non-monotonic behavior, initially decreasing from 16.14±0.04 × 10$^{-6}$ K$^{-1}$ for the homogenized sample to a minimum of 15.32±0.05 ×10$^{-6}$ K$^{-1}$ at intermediate strains ($N$ = 1/4 turn) before increasing to 16.23±0.05 × 10$^{-6}$ MK$^{-1}$ with further straining up to $N$ = 4 turns. The possible explanation for these changes is the contribution of defect-induced residual stress and grain boundaries, as will be discussed in the next section [53].

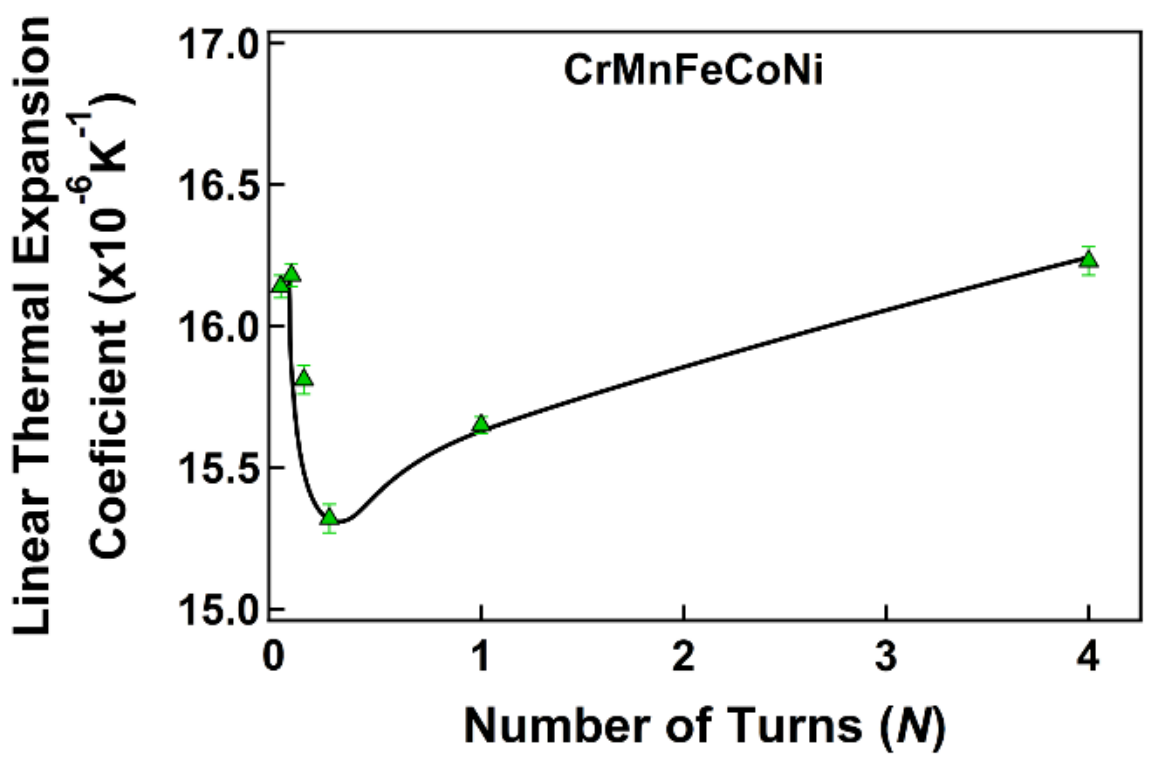

Fig. 8. Initial decrease followed by an increase in thermal expansion of high-entropy alloy CrMnFeCoNi treated via severe plastic deformation. Linear thermal expansion coefficients obtained from synchrotron analysis versus number of HPT rotations.

## 4. Discussion

The current study extends prior applications of SPD to HEAs, focused mainly on enhancing mechanical properties [9-12] and partly on functional characteristics [18-26], to the tailoring of electrical and thermal properties. The findings of this study can be strategically employed to tailor transport properties in HEAs for applications requiring combined high strength and controlled thermal/electrical performance, such as high-power electrical contacts, refractory materials for fusion reactor components or thermoelectric devices [28-30]. The anomalous increase in electrical conductivity of CrMnFeCoNi from the coarse-grained homogenized state to the nanograined state after HPT treatment for $N = 4$ rotations reveals a fundamental coupling of electronic and phononic transport mechanisms [13,15-17], an issue that needs to be discussed. Moreover, changes in thermal expansion need further discussion.

### 4.1. Phonon Scattering and Thermal Conductivity

The thermal conductivity of the HEA can be quantitatively analyzed by comprising electronic ($\kappa_e$) and phononic ($\kappa_p$) contributions [54].

$$\kappa = \kappa_e + \kappa_p \quad (3)$$

Using the Wiedemann-Franz law with the Sommerfeld Lorenz number $L_0 = 2.44\times10^{-8}$ WΩK$^{-2}$, $\kappa_e$ can be calculated from electrical resistivity ($\rho_e = 1/\sigma$, $\sigma$: electrical conductivity) for a given temperature ($T$) [54].

$$\kappa_e = L_0 T/\rho_e \quad (4)$$

The contribution of electronic and phononic scattering to transport is shown in Fig. 9a, where electrical and thermal data were compared at close temperatures. For the homogenized condition at room temperature, calculations lead to $\kappa_e = 6.97$ W m$^{-1}$ K$^{-1}$, yielding a phononic contribution of $\kappa_p = 3.23$ W m$^{-1}$ K$^{-1}$. After HPT for $N = 4$, the electronic conductivity increases slightly and reaches $\kappa_e = 7.34$ W m$^{-1}$ K$^{-1}$, while phononic conductivity increases to 6.56 W m$^{-1}$ K$^{-1}$. The Sommerfeld Lorenz number $L_0$ is the standard degenerate-limit value and has been widely applied to CrMnFeCoNi and related HEAs [48]. Because the phonon contribution is obtained by subtraction, $\kappa_p = \kappa_{total} - L_0T/\rho_e$, any reduction of $L$ below $L_0$ would increase the inferred $\kappa_p$ and therefore reinforce the conclusion that phonon transport is substantially enhanced by HPT. The 100% increase in $\kappa_p$ demonstrates that, despite expected phonon scattering at defects, an increase in heat capacity and an unusual increase in phonon mean free path dominate the thermal transport enhancement of the HEA by HPT processing. The phonon mean free path ($\Lambda_p$) can be determined from kinetic theory [54].

$$\kappa_p = C_{P}\rho v_s \Lambda_p/3 \qquad (5)$$

where $C_P$ is the specific heat capacity (per mass unit under constant pressure). $\rho$ is density, and $v_s$ is the sound velocity (4416 m s$^{-1}$ for CrMnFeCoNi [55]). The phonon mean free path is calculated as 0.6 nm for the homogenized sample, but it increases to 1.1 nm after HPT processing for $N = 4$, as shown in Fig. 9b. Although the kinetic model yields such values, this length scale is comparable to interatomic distances, indicating that the assumption of well-defined propagating phonons is no longer valid. In this regime, the Ioffe–Regel criterion is approached [56], where the phonon mean free path becomes comparable to the phonon wavelength, and heat transport is dominated by strongly scattered or non-propagating vibrational modes rather than ballistic phonons [54].

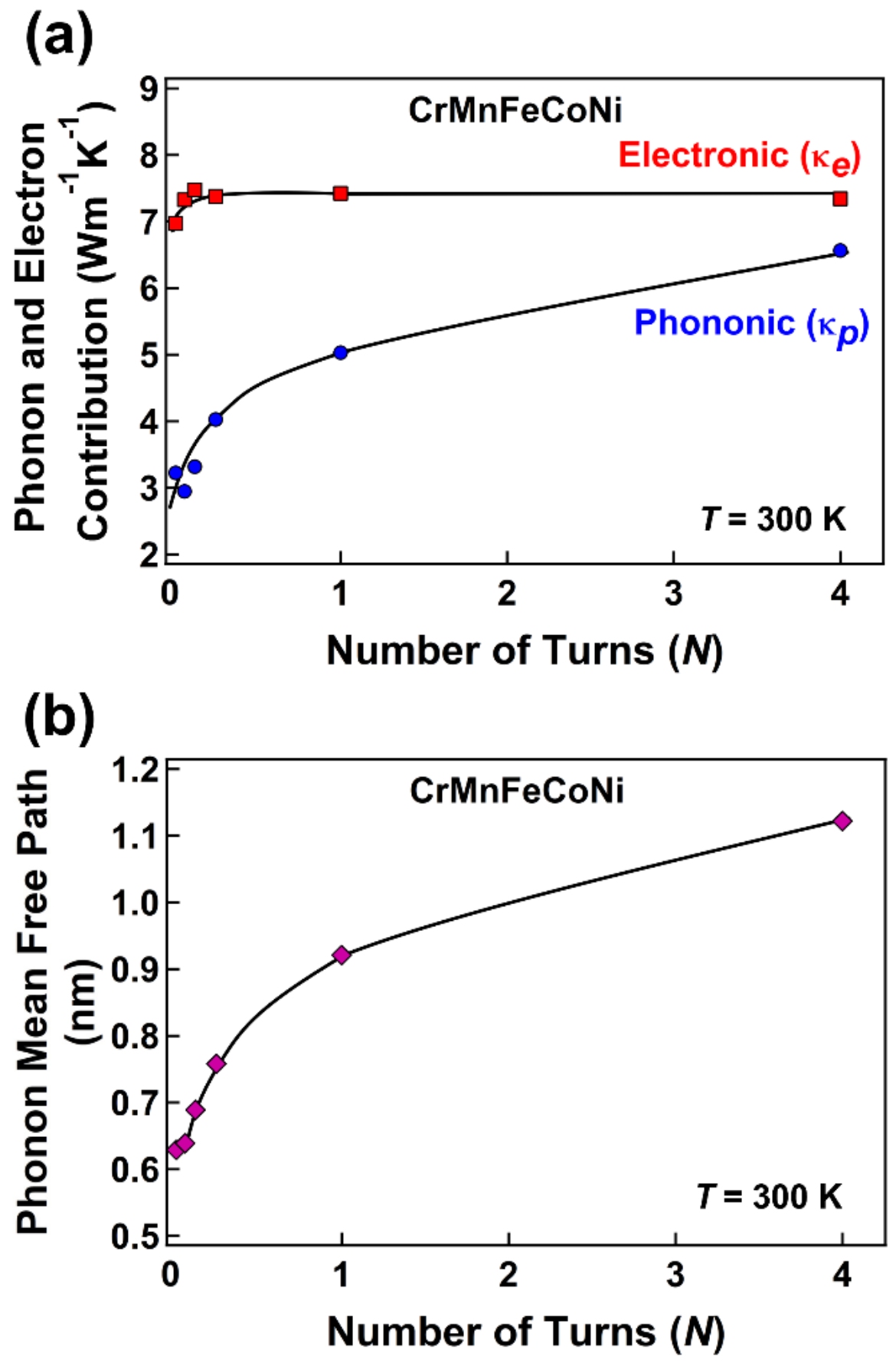

Fig. 9. Increasing contribution of electronic phenomena by 5% and phononic phenomena by 100% to unusual electrical and thermal conductivity of high-entropy alloy CrMnFeCoNi treated via severe plastic deformation with phonon mean free path at the interatomic distance level. Variations of (a) electronic and phononic contributions to conductivity and (b) photon mean free path versus number of HPT rotations.

To further evaluate the phonon transport, the minimum thermal conductivity, $\kappa_{min}$, was estimated using the Cahill-Pohl model [33].

$$\kappa_{min} = (\pi/6)^{1/3} k_B n^{2/3} (v_L + 2v_T) \quad (6)$$

where $k_B$ is the Boltzmann constant, $n$ is the atomic number density, and $v_L$ and $v_T$ are the longitudinal and transverse sound velocities, respectively. The atomic number density is estimated as $8.5\times10^{28}$ m$^{-3}$ using the measured density (7904 kg m$^{-3}$) and the equiatomic average molar mass. The sound velocities can be approximated using standard relations for isotropic solids, where $v_L$ = 1.7 $v_T$ and $v_T$ = 0.6 $v_S$ are based on typical elastic constant ratios in metals [54]. Using $v_S$ = 4416 m s$^{-1}$ yields $v_T$ = 2650 m s$^{-1}$ and $v_L$ = 4500 m s$^{-1}$. Substituting these values into the above equation gives $\kappa_{min}$ = 2.11 W m$^{-1}$ K$^{-1}$. The phonon contribution of the homogenized alloy (3.23 W m$^{-1}$ K$^{-1}$) is therefore within a factor of ~1.5 of $\kappa_{min}$, indicating that heat transport in the coarse-grained state is close to the minimum thermal conductivity limit. After HPT for $N$ = 4, however, $\kappa_p$ increases to 6.56 W m$^{-1}$ K$^{-1}$, i.e. about three times $\kappa_{min}$, showing that the deformation moves phonon transport away from the strongly scattered minimum-conductivity regime. Therefore, the observed increase in phononic thermal conductivity after HPT cannot be attributed to reduced scattering at grain boundaries or dislocations, since these microstructural length scales are much larger than the effective transport length. Instead, it is hypothesized that the enhancement is more plausibly associated with modifications in atomic-scale disorder, such as deformation-induced short-range ordering, which can reduce phonon scattering and increase the diffusivity of vibrational modes in high-entropy alloys [57,58]. Direct experimental verification of short-range ordering by HPT processing would further clarify the underlying mechanism.

### 4.2. Electron Scattering and Electrical Conductivity

Regarding the increase in electrical conductivity, it is important to first refer to conventional defect scattering theory [48].

$$\rho_e = \rho_0 + \rho_{gb} + \rho_{dis} \tag{7}$$

Here, $\rho_0$ is the residual inherent resistivity, which depends on the composition, chemical disorder and point defects, $\rho_{gb}$ is the grain boundary scattering contribution, which is dependent on the grain size features and fraction, and $\rho_{dis}$ is the dislocation scattering contribution, which depends on the dislocation density. Since densities of dislocations and grain boundaries both increase after HPT processing, the process should lead to a significant resistivity increase [14,46,47], which contradicts experimental measurements in this study. The anomalous conductivity enhancement suggests two major possible reasons. First, the intrinsic chemical disorder in HEAs creates an exceptionally high inherent resistivity ($\rho_0$ = 10$^{-7}$ Ωm for the homogenized sample with large grains and low dislocation density), meaning additional defect scattering has a proportionally smaller impact compared to inherent resistivity [58]. Second, HPT processing may generate deformation-induced short-range ordering (SRO) [59] that reduces the intrinsic lattice distortion scattering term $\rho_0$. Recent first-principles calculations combined with experiments showed SRO can decrease $\rho_0$ in multi-component alloys by reducing electron localization [60,61]. It should be noted that diffusion of constituent elements is expected to drive changes in local structure and chemical composition fluctuations in HEAs to generate SRO [62], an effect that was experimentally observed at very high temperatures [63]. SPD involves the generation of vacancies, dislocations and non-equilibrium grain boundaries that greatly reduce energy barriers for diffusion, allowing the occurrence of high-temperature phenomena at ambient temperature [64,65].

The electrical resistivity decrease in this study is consistent with Ref. [31], where a similar effect in cold-rolled CrMnFeCoNi was attributed to the destruction of mesoscopic chemically modulated domains. In Equation (7), such destruction would reduce the intrinsic disorder-scattering term $\rho_0$, competing with increases in $\rho_{gb}$ and $\rho_{dis}$. The current data do not distinguish this mechanism from deformation-induced short-range ordering; both remain plausible. The thermal data independently indicate that the conductivity increase is not governed by conventional defect scattering, since the phonon contribution rises while the phonon mean free path remains near interatomic distances. Understanding the exact reason for unusual electron scattering in HEAs and its connection with SRO needs future theoretical studies and atomic-scale simulations.

### 4.3. Defect Relaxation and Thermal Expansion

Regarding the linear thermal expansion, the initial decrease in thermal expansion is associated with the accumulation of lattice defects and the development of internal stresses during early stages of deformation. The high density of dislocations and dislocation locks (e.g. Lomer-Cottrell locks) introduces significant stored energy and long-range internal stress fields, which can constrain lattice response and contribute to stress relaxation processes during heating, thereby reducing the apparent thermal expansion coefficient [49,50]. The thermal expansion coefficient, $\alpha$, can be expressed by the following relationship [54].

$$\alpha=\gamma_G C_V/(KV) \tag{8}$$

where $\gamma_G$ is the Grüneisen parameter reflecting anharmonic lattice vibrations, $C_V$ is specific heat capacity (per mass unit under constant volume), $K$ is the bulk modulus, and $V$ is the volume. In severely deformed materials, the measured thermal expansion includes both intrinsic lattice expansion and non-equilibrium contributions associated with defect recovery, particularly point defects with a lower thermal stability [66,67], and relevant stress relaxation.

$$\alpha=\alpha_{\text{Lattice}}+\alpha_{\text{Relaxtaion}} \tag{9}$$

As a result, the apparent thermal expansion can be reduced in the presence of a high density of defects. After reaching a minimum at intermediate strain, the subsequent increase in thermal expansion correlates with the evolution toward a nanograined structure containing a high fraction of grain boundaries, which are thermally more stable [66,67]. These boundaries are characterized by excess free volume and enhanced anharmonicity compared to grain interiors, which promote larger thermal expansion through an increase in the Grüneisen parameter and a reduction in effective stiffness [53]. This transition from a defect- and stress-dominated regime to a grain-boundary- and free-volume-dominated regime highlights the interplay of competing mechanisms governing the thermophysical response of severely deformed HEAs.

## 5. **Conclusions**

In summary, processing CrMnFeCoNi by high-pressure torsion (HPT) leads to increasing hardness, electrical conductivity, specific heat capacity, and thermal conductivity, while the linear thermal expansion coefficient exhibits a non-monotonic behavior, initially decreasing and then increasing with strain. The increase in thermal conductivity in HPT-processed CrMnFeCoNi is primarily associated with reduced phonon scattering, with the phonon contribution increasing by ~100%, whereas the electronic contribution increases by only ~5%. The phonon mean free path remains on the order of interatomic distances, and comparison with the Cahill-Pohl model indicates that heat transport occurs near the minimum thermal conductivity limit. These results suggest that the enhancement in thermal transport is not governed by conventional defect scattering, but it is hypothesized that modifications in atomic-scale disorder and short-range ordering are responsible for these unusual observations. The increase in electrical conductivity further indicates non-trivial electron–defect interactions in this chemically complex system. Overall, these findings provide new insights into defect engineering and its role in tuning electrical and thermal transport properties of functional high-entropy alloys.

## CRediT Authorship Contribution Statement

All authors: Conceptualization, Methodology, Investigation, Validation, Writing – review & editing.

## Declaration of competing interest

The authors declare no competing interests that could have affected the presentation in this manuscript.

**Acknowledgements**

This research was supported partly by the Japan Science and Technology Agency (JST) through the program Establishment of University Fellowships Towards the Creation of Science Technology Innovation (JPMJFS2132) and ASPIRE (JPMJAP2332).

**Data Availability**

The data will be available for sharing upon contact with the corresponding author.

**References**

[1] K. Biswas, J.W. Yeh, P.P. Bhattacharjee, J.T.M. DeHosson, High entropy alloys: key issues under passionate debate, Scr. Mater. 188 (2020) 54-58.

[2] B. Cantor, Multicomponent high-entropy Cantor alloys, Prog. Mater. Sci. 120 (2021) 100754.

[3] D.B. Miracle, O.N. Senkov, A critical review of high entropy alloys and related concepts, Acta Mater. 122 (2017) 448-511.

[4] E.P. George, D. Raabe, R.O. Ritchie, High-entropy alloys, Nat. Rev. Mater. 4 (2019) 515-534.

[5] R.Z. Valiev, Y. Estrin, Z. Horita, T.G. Langdon, M.J. Zehetbauer, Y. Zhu, Producing bulk ultrafine-grained materials by severe plastic deformation: ten years later, JOM 68 (2016) 1216-1226.

[6] K. Edalati, A.Q. Ahmed, S. Akrami, K. Ameyama, V. Aptukov, R.N. Asfandiyarov, M. Ashida, V. Astanin, A. Bachmaier, V. Beloshenko, E.V. Bobruk, K. Bryła, J.M. Cabrera, A.P. Carvalho, N.Q. Chinh, I.C. Choi, R. Chulist, J.M. Cubero-Sesin, G. Davdian, M. Demirtas, S. Divinski, K. Durst, J. Dvorak, P. Edalati, S. Emura, N.A. Enikeev, G. Faraji, R.B. Figueiredo, R. Floriano, M. Fouladvind, D. Fruchart, M. Fuji, H. Fujiwara, M. Gajdics, D. Gheorghe, Ł. Gondek, J.E. González-Hernández, A. Gornakova, T. Grosdidier, J. Gubicza, D. Gunderov, L. He, O.F. Higuera, S. Hirosawa, A. Hohenwarter, Z. Horita, J. Horky, Y. Huang, J. Huot, Y. Ikoma, T. Ishihara, Y. Ivanisenko, J.I. Jang, A.M. Jorge Jr, M. Kawabata-Ota, M. Kawasaki, T. Khelfa, J. Kobayashi, L. Kommel, A. Korneva, P. Kral, N. Kudriashova, S. Kuramoto, T.G. Langdon, D.H. Lee, V.I. Levitas, C. Li, H.W. Li, Y. Li, Z. Li, H.J. Lin, K.D. Liss, Y. Liu, D.M. Marulanda Cardona, K. Matsuda, A. Mazilkin, Y. Mine, H.

Miyamoto, S.C. Moon, T. Müller, J.A. Muñoz, M.Y. Murashkin, M. Naeem, M. Novelli, D. Olasz, R. Pippan, V.V. Popov, E.N. Popova, G. Purcek, P. de Rango, O. Renk, D. Retraint, Á. Révész, V. Roche, P. Rodriguez-Calvillo, L. Romero-Resendiz, X. Sauvage, T. Sawaguchi, H. Sena, H. Shahmir, X. Shi, V. Sklenicka, W. Skrotzki, N. Skryabina, F. Staab, B. Straumal, Z. Sun, M. Szczerba, Y. Takizawa, Y. Tang, R.Z. Valiev, A. Vozniak, A. Voznyak, B. Wang, J.T. Wang, G. Wilde, F. Zhang, M. Zhang, P. Zhang, J. Zhou, X. Zhu, Y.T. Zhu, Severe plastic deformation for producing superfunctional ultrafine-grained and heterostructured materials: an interdisciplinary review, J. Alloys Compd. 1002 (2024) 174667. https://doi.org/10.1016/j.jallcom.2024.174667.

[7] A.P. Zhilyaev, T.G. Langdon, Using high-pressure torsion for metal processing: Fundamentals and applications, Prog. Mater. Sci. 53 (2008) 893–979.

[8] K. Edalati, Z. Horita, A review on high-pressure torsion (HPT) from 1935 to 1988, Mater. Sci. Eng. A 652 (2016) 325–352.

[9] J. Gubicza, P.T. Hung, Nanostructuring of multi-principal element alloys by severe plastic deformation: from fundamentals to an improved functionality, Mater. Trans. 64 (2023) 1284-1298.

[10] D.H. Lee, I.C. Choi, M. Kawasaki, T.G. Langdon, J.I. Jang, A review of recent research on nanoindentation of high-entropy alloys processed by high-pressure torsion, Mater. Trans. 64 (2023) 1551–1565.

[11] H. Shahmir, M.S. Mehranpour, M. Kawasaki, T.G. Langdon, Superplasticity in severely deformed high-entropy alloys, Mater. Trans. 64 (2023) 1526-1536.

[12] W. Skrotzki, R. Chulist: Severe plastic deformation of high-entropy alloys, Mater. Trans. 64 (2023) 1769–1783.

[13] Q. G. Zhang, B. Y. Cao, X. Zhang, M. Fuji, K. Takahashi, Influence of grain boundary scattering on the electrical and thermal conductivities of polycrystalline gold nanofilms, Phys. Rev. B 74 (2006) 134109.

[14] K. Edalati, K. Imamura, T. Kiss, Z. Horita, Equal-channel angular pressing and high-pressure torsion of pure copper: evolution of electrical conductivity and hardness with strain, Mater. Trans. 53 (2012) 123-127.

[15] A. Giri, P. E. Hopkins, A review of experimental and computational advances in thermal boundary conductance and nanoscale thermal transport across solid interfaces, Adv. Funct. Mater. 30 (2020) 1903857.

[16] Y. Dong, S. Mu, X. Guo, J. Han, J. Duan, N. Jia, Y. Xue, K. Jin, Transport properties of refractory high-entropy alloys with single-phase body-centered cubic structure, Scr. Mater. 231 (2023) 115464.

[17] H. Wang, S. Ma, W. Zhao, Q. He, Y. Liu, S. Zhao, Y. Yang, Exceptionally low thermal conductivity in distorted high entropy alloy, Mater. Res. Lett. 13 (2025) 24-34.

[18] P. Edalati, M. Fuji, K. Edalati, Superfunctional high-entropy alloys and ceramics by severe plastic deformation, Rare Met. 42 (2023) 3246-3268.

[19] A. Mohammadi, P. Edalati, M. Arita, J.W. Bae, H.S. Kim, K. Edalati, Microstructure and defect effects on strength and hydrogen embrittlement of high-entropy alloy CrMnFeCoNi processed by high-pressure torsion, Mater. Sci. Eng. 844 (2022) 143179.

[20] K. Edalati, A.M. Bertrand, P. Edalati, T.T. Nguyen, N. Enikeev, M. Mito, Influence of nanostructuring through high-pressure torsion (HPT) on superconductivity of a high-entropy alloy, Scr. Mater. 266 (2025) 116769.

[21] P. Edalati, A. Cremasco, K. Edalati, R. Floriano, High corrosion resistance of nanograined and nanotwinned $Al_{0.1}CoCrFeNi$ high-entropy alloy processed by high-pressure torsion, Intermetallics 172 (2024) 108368.

[22] T.T. Nguyen, P. Edalati, S. Dangwal, K.D. Pereira, A. Cremasco, R. Floriano, A.D. Luchessi, K. Edalati, Boosting biocompatibility and mechanical property evolution in a high-entropy alloy via nanostructure engineering and phase transformations, J. Alloys Compd. 1035 (2025) 181438.

[23] J. Hidalgo-Jiménez, J.M. Cubero-Sesín, K. Edalati, S. Khajavi; J. Huot, Effect of high-pressure torsion on first hydrogenation of Laves phase $Ti_{0.5}Zr_{0.5}(Mn_{1-x}Fe_x)Cr_1$ (x = 0, 0.2 and 0.4) high entropy alloy, J. Alloys Compd. 969 (2023) 172243.

[24] H.T.N. Hai, M. Arita, K. Edalati, High-entropy perovskites as new photocatalysts for cocatalyst-free water splitting, Appl. Catal. B 383 (2026) 126081.

[25] J. Hidalgo-Jiménez, T. Akbay, T. Ishihara, K. Edalati, Active high-entropy photocatalyst designed by incorporating alkali metals to achieve $d^0$+$d^{10}$+$s^0$ cationic configurations and wide electronegativity mismatch, Acta Mater. 32 (2026) 121632.

[26] T.T. Nguyen, K. Edalati, High-entropy oxide with tailored heterogeneous electronic structure as a low-bandgap catalyst for antibiotic photodegradation under visible light, Appl. Catal. B 382 (2026) 126011.

[27] Q. Tang, B. Jiang, K. Wang, W. Wang, B. Jia, T. Ding, Z. Huang, Y. Lin, J. He, High-entropy thermoelectric materials, Jule 8 (2024) 1641-1666.

[28] P. Sobota, R. Topolnicki, T. Ossowski, T. Pikula, A. Pikul, R. Idczak, Superconductivity in the high-entropy alloy (NbTa)0.67(MoHfW)0.33, Phys. Rev. B 106 (2022) 184512.

[29] Y. Cui, J. Shen, S.M. Manladan, K. Geng, S. Hu, Wear resistance of FeCoCrNiMnAlx high-entropy alloy coatings at high temperature, Appl. Surf. Sci. 512 (2020) 145736.

[30] S. Dangwal, K. Edalati, Developing a single-phase and nanograined refractory high-entropy alloy ZrHfNbTaW with ultrahigh hardness by phase transformation via high-pressure torsion, J. Alloys Compd. 1010 (2025) 178274.

[31] Y. Miyajima, T. Nagata, K. Takeda, S. Yoshida, S. Yasuno, C. Watanabe, K. Ishikawa, H. Adachi, N. Tsuji, Destruction of mesoscopic chemically modulated domains in single phase high entropy alloy via plastic deformation, Sci. Rep. 12 (2022) 16776.

[32] J. Hidalgo-Jiménez, P. Edalati, M.A. Islam, M. Arita, B.B. Saha, K. Edalati, Thermal and electrical conductivity of a refractory high-entropy alloy after high-pressure torsion: Electron versus phonon contributions, J. Phys. Chem. Solids 219 (2026) 114013.

[33] D.G. Cahill, P.V. Braun, G. Chen, D.R. Clarke, S.i Fan, K.E. Goodson, P. Keblinski, W.P. King, G.D. Mahan, A. Majumdar, H.J. Maris, S.R. Phillpot, E. Pop, L.S. Crossmark, Nanoscale thermal transport. II. 2003–2012, Appl. Phys. Rev. 1 (2014) 011305.

[34] S. Haas, M. Mosbacher, O.N. Senkov, M. Feuerbacher, J. Freudenberger, S. Gezgin, R. Völkl, U. Glatzel, Entropy determination of single-phase high entropy alloys with different crystal structures over a wide temperature range, Entropy 20 (2018) 654.

[35] V. Dyadkin, P. Pattison, V. Dmitriev, D. Chernyshov, A new multipurpose diffractometer PILATUS@SNBL, J. Synchrotron Rad. 23 (2016) 825-829.

[36] G. Ashiotis, A. Deschildre, Z. Nawaz, J.P. Wright, D. Karkoulis, F.E. Picca, J. Kieffer, The fast azimuthal integration Python library: pyFAI, J. Appl. Crystallogr. 48 (2015) 510-519.

[37] J. Rodríguez-Carvajal, Recent advances in magnetic structure determination by neutron powder diffraction, Physica B 192 (1993) 55-69.

[38] M.J. Cliffe and A.L. Goodwin, J. Appl. Cryst. (2012). 45, 1321-1329.

[39] B. Wei, W. Wu, M. Gong, S. Yu, S. Ni, M. Song, J. Wang, Influence of lowering basal stacking fault energy on twinning behaviours, Acta Mater. 245 (2023) 118637.

[40] S.H. Joo, H. Kato, M.J. Jang, J. Moon, C.W. Tsai, J.W. Yeh, H.S. Kim, Tensile deformation behavior and deformation twinning of an equimolar CoCrFeMnNi high-entropy alloy, Mater. Sci. Eng. A 689 (2017) 122-123.

[41] J.C.M. Li, C.R. Feng, B.B. Rath, Emission of dislocations from grain boundaries and its role in nanomaterials, Crystals 11 (2021) 41.

[42] T. Shimokawa, T. Niiyama, T. Miyaki, M. Ikeda, K. Higashida, A novel work hardening mechanism of nanoscale materials by grain boundary transformation, Acta Mater. 224 (2022) 117536.

[43] A. Mohammadi, M. Novelli, M. Arita, J.W. Bae, H.S. Kim, T. Grosdidier, K. Edalati, Gradient-structured high-entropy alloy with improved combination of strength and hydrogen embrittlement resistance, Corros. Sci. 200 (2022) 110253.

[44] S.H. Shim, M. Saboktakin Rizi, H. Minouei, S.I. Hong, Role of Cu in nanostructural relationship between phase separation and deformation-induced twinning in heavily drawn non-equiatomic high-entropy alloy wire, Nanomaterials 15 (2025) 1281.

[45] H. Minouei, M. Saboktakin Rizi, M. Ebrahimian, S.B. Akçay, S. Heo, T. Varol, O. Güler, N. Park, S.I. Hong, Severe plastic deformation of high-entropy alloys: A critical review of deformation mechanisms and microstructural evolution, J. Alloys Compd. 1079 (2026) 189852.

[46] J.E. González-Hernández, J.M. Cubero-Sesin, Electrical conductivity of ultrafine-grained cu and al alloys: attaining the best compromise with mechanical properties, Mater. Trans. 64 (2023) 1754-1768.

[47] M.Y. Murashkin, N.A. Enikeev, X. Sauvage, Potency of severe plastic deformation processes for optimizing combinations of strength and electrical conductivity of lightweight Al-based conductor alloys, Mater. Trans. 64 (2023) 1833-1843.

[48] A. Berger, S. Benito, A. Konchits, G. Laplanche, B. Shanina, S. Weber, Thermophysical properties of equiatomic CrMnFeCoNi, CrFeCoNi, CrCoNi, and CrFeNi high- and medium-entropy alloys, Mater. Today Commun. 39 (2024) 109341.

[49] J. Bevk, The effect of lattice defects on the low-temperature heat capacity of copper, Philos. Mag. 28 (1973) 1379-1390.

[50] P.C. Schuck, J. Marian, J.B. Adams, B. Sadigh, Vibrational properties of straight dislocations in bcc and fcc metals within the harmonic approximation, Philos. Mag. 89 (2009) 2861–2882.
[51] J.R. Wang, D.H. Zhu, J.L. Chen, B.C. Pan, Effect of lattice defects, hydrogen impurity and temperature on electronic thermal conductivity in first wall tungsten materials, J. Nucl. Mater. 581 (2023) 154418.
[52] X. Qian, J. Zhou, G. Chen, Phonon-engineered extreme thermal conductivity materials, Nat. Mater. 20 (2021) 1188-1202.
[53] R. Daniel, D. Holec, M. Bartosik, J. Keckes, C. Mitterer, Size effect of thermal expansion and thermal/intrinsic stresses in nanostructured thin films: Experiment and model, Acta Mater. 59 (2011) 6631–6645.
[54] J.M. Ziman, Electrons and Phonons, Oxford University Press, 2001.
[55] R. Nikbakht, M. Saadati, T.S. Kim, M. Jahazi, H.S. Kim, B. Jodoin, Cold spray deposition characteristic and bonding of CrMnCoFeNi high entropy alloy, Surf. Coat. Technol. 425 (2021) 127748.
[56] A.F. Ioffe, A.R. Regel, Non-crystalline, amorphous, and liquid electronic semiconductors, Prog. Semicond. 4 (1960) 237-291.
[57] Y. Ma, Q. Wang, C. Li, L.J. Santodonato, M. Feygenson, C. Dong, P.K. Liaw, Chemical short-range orders and the induced structural transition in high-entropy alloys, Scr. Mater. 144 (2018) 64-68.
[58] K. Jin, B.C. Sales, G.M. Stocks, G.D. Samolyuk, M. Daene, W.J. Weber, Y. Zhang, H. Bei, Tailoring the physical properties of Ni-based single-phase equiatomic alloys by modifying the chemical complexity, Sci. Rep. 6 (2016) 20159.
[59] J.B. Seol, W.S. Ko, S.S. Sohn, M.Y. Na, H.J. Chang, Y.U. Heo, J.G. Kim, H. Sung, Z. Li, E. Pereloma, H.S. Kim, Mechanically derived short-range order and its impact on the multi-principal-element alloys, Nat. Commun. 13 (2022) 6766.
[60] C. Niu, A.J. Zaddach, A.A. Oni, X. Sang, J.W. Hurt, III, J.M. LeBeau, C.C. Koch, D.L. Irving, Spin-driven ordering of Cr in the equiatomic high entropy alloy NiFeCrCo, Appl. Phys. Lett. 106 (2015) 161906.
[61] Z. Wang, C.T. Liu, P. Dou, Thermodynamics of vacancies and clusters in high-entropy alloys, Phys. Rev. Mater. 1 (2017) 043601.

[62] J. Zhang, Z.-Y. Ma, K.-H. Xun, Z.-D. Sha, J. Ding, Local chemical order in CrFeCoNiPd high entropy alloys, J. Appl. Phys. 138 (2025) 045101.

[63] A. Smekhova, D. Gaertner, A. Kuzmin, A.G. Buzanich, G. Schuck, I. Zizak, G. Wilde, K.V. Yusenko, S. Divinski, Anomalies in the short-range local environment and atomic diffusion in single crystalline equiatomic CrMnFeCoNi high-entropy alloy, Nano Res. 17 (2024) 5336–5348.

[64] B.B. Straumal, A.A. Mazilkin, B. Baretzky, G. Schütz, E. Rabkin, R.Z. Valiev, Accelerated diffusion and phase transformations in Co–Cu alloys driven by the severe plastic deformation, Mater. Trans. 53 (2012) 63–71.

[65] G. Wilde, S. Divinski, Grain boundaries and diffusion phenomena in severely deformed materials, Mater. Trans. 60 (2019) 1302–1315.

[66] D. Setman, E. Schafler, E. Korznikova, M.J. Zehetbauer, The presence and nature of vacancy type defects in nanometals detained by severe plastic deformation, Mater. Sci. Eng. A 493 (2008) 116-122.

[67] B. Oberdorfer, B. Lorenzoni, K. Unger, W. Sprengel, M. Zehetbauer, R. Pippan, R. Würschum, Absolute concentration of free volume-type defects in ultrafine-grained Fe prepared by high-pressure torsion, Scr. Mater. 63 (2010) 452-455.